\documentclass{aa}
\usepackage{graphicx,times}
\begin{document}

\title{Measuring the Absolute Height and Profile of the Mesospheric
Sodium Layer using a Continuous Wave Laser}
\titlerunning{Measuring the Mesospheric Sodium Layer using a
Continuous Wave Laser}

\author{D. J. Butler\inst{1}  \and R. I. Davies\inst{2}
\and R. M. Redfern\inst{3} \and N. Ageorges\inst{4}  \and
H. Fews\inst{3}\thanks{Current address: Marconi Communications, ONDATA GmbH, Germany} %\and S. Hippler (?)\inst{1}  
}

%\offprints{D. Butler}

\institute{Max-Planck-Institut f\"ur Astronomie, K\"onigstuhl 17,
D-69117 Heidelberg, Germany \\
\email{butler@mpia.de}
\and
Max-Planck-Institut f\"ur extraterrestrische Physik, Postfach 1312, D-85741 Garching, Germany
%\email{davies@mpe.mpg.de}
\and
Experimental Physics Department, National University of Ireland, Galway, Ireland
%\email{redfern@physics.nuigalway.ie}
\and
European Southern Observatory, Alonso de C\'ordova 3107, Vitacura, Casilla
19001, Santiago 19, Chile
%\email{nageorge@eso.org}
%\and
%Experimental Physics Department, National University of Ireland, Galway,
%Ireland\thanks{Current address: }
%\email{hfews@....}         
}
\authorrunning{Butler et al.}

\date{Received ...; accepted ...}

\abstract{
We have developed and tested a novel method, based on LIDAR, of
measuring the height and profile of the mesospheric 
sodium layer using a continuous wave laser.
It is more efficient than classical LIDAR as the laser is on for 50\%
of the time, and so can in principle be used during laser guide star
adaptive optics observations.
It also has significant advantages over direct imaging techniques
because it does not require a second telescope, is almost independent of the
atmospheric conditions, and avoids triangulation problems in
determining the height.
In the long term, regular monitoring using this method would allow
a valuable database of sodium layer profiles, heights, and return flux
measurements to be built up 
 which would enable  observatory staff astronomers to schedule
observations optimally.
In this paper we describe the original experiment carried out using
the ALFA laser guide star system at Calar Alto Observatory in Spain.
We validate the method by comparing the LIDAR results with those
obtained from simultaneous imaging from an auxiliary
telescope.
Models are presented of a similar system to be implemented in the {\em
Very Large Telescope} Laser Guide Star Facility, which will enable the
initial focus setting for the adaptive optics systems to be determined
with an accuracy of less than 200\,m on a timescale of 1\,minute.


}       %end of abstract

\maketitle

%________________________________________________________________

\section{Introduction}
\label{intro}

It is well known in the astronomical community 
that classical astronomical adaptive optics (AO) systems are
only able to achieve their best performance over a few percent of the
sky due to the scarcity of sufficiently bright reference (or guide) stars.
Fortunately this severe limitation can be overcome by the use of a
laser guide star (LGS), as suggested by Foy \& Labeyrie (\cite{foy85}).
The principal advantage of a LGS is that it produces a bright reference
source that can be pointed almost anywhere on the sky.
Although a tip-tilt star is still needed, it can be further from the science
target and much fainter, vastly increasing the sky accessible to
adaptive optics systems.
The main disadvantage of a LGS is that its proximity   to the telescope
means that it samples less of the atmospheric column seen by a natural
star (the `cone effect', or `focal anisoplanaticism') resulting in
poorer wavefront correction.
For 8.0-m class telescopes which plan to attain the best
correction possible with only a single LGS, the only
reasonable choice is a sodium LGS generated by resonant excitation of
 sodium atoms in the Earth's atmosphere.
% naturally occurring sodium atoms.
In both the northern and southern hemispheres these are found only in
the mesosphere at a height of $\sim$90\,km (see She et
al. \cite{she00}, and references therein).
Atomic sodium abundance has been observed to vary seasonally 
by a factor of 3--4, but is typically between about 
$2\times10^{13}$\,m$^{-2}$ around the solstices
and 4--$5\times10^{13}$\,m$^{-2}$ around the equinoxes at
mid-latitudes (see Table~\ref{min_abundances}).

Most of the sodium lasers which are planned for major astronomical
observatories are designed to compensate for this so that
they can be used at any time of year.
For example, the {\em Very Large Telescope} (VLT) Laser Guide Star
Facility (LGSF; Bonaccini et al. \cite{bon02}) will provide a minimum of
10$^6$\,photons\,m$^{-2}$\,s$^{-1}$ (equivalent to $\sim$1000\,photons per
centroid measurement in a Shack-Hartmann wavefront sensor) 
even with a sodium column density of
2$\times$10$^{13}$\,m$^{-2}$ during median atmospheric conditions.
The launched laser power must then be at least 6\,W (continuous wave,
with four single frequency lines equally spaced within a 500\,MHz
envelope).
The laser used at the Keck Observatory is designed to meet similar
requirements, and  has a launch power of 15--20\,W (pulsed,
broadband spectrum; Pennington et al. \cite{pen02}).

The occurrence of mesospheric atomic sodium is a natural phenomenon. 
Meteoric ablation in the upper atmosphere is thought to be the major 
contributor, with sudden short-lived increases of sodium density
in localized regions of the layer known to occur (see 
Collins et al. \cite{col02} and references therein for the various
proposed explanations).
These sporadic layers typically arise
in thin layers about 1--3\,km wide at altitudes between 90 and 105\,km.
Since the increase in atomic density can be an order of magnitude
more than the background mesospheric sodium density, a shift in
the mean layer altitude could occur. 
Shifts as large as 400\,m on timescales of  1--2\,mins have been observed by O'Sullivan et al. (\cite{sul00}). 
Such shifts would cause a LGS defocus (as well as, to a lesser extent,
astigmatism and other higher order wavefront aberrations) 
to be sensed by a wavefront sensor.
The rms wavefront error arising from uncorrected LGS defocus at a
telescope with pupil size $D$ is given by
$\sigma_\lambda \, = D^2 \, \delta H \, / \, 16 \, \sqrt{3} \, H^2$ 
(Louarn et al. \cite{lou00}) where $H$ is the height of the sodium
layer ($\sim$90\,km). 
The corresponding Strehl loss is 
$S_{\rm loss} = 1 - \exp{(-(2\,\pi\,\sigma_\lambda\,/\,\lambda)^2)}$ 
where $\lambda$ is the  wavelength
  at which improved angular resolution is required. 
For the 3.5-m at Calar Alto, where the measurements above were taken,
a LGS defocus of $\delta H = 400$\,m would introduce a wavefront
error of $\sigma_\lambda = 22$\,nm and a Strehl loss of $<0.5$\% at an
observing wavelength of 2.2\,$\mu$m.
However, for an 8.0-m telescope we find  $\sigma_\lambda = 110$\,nm
and the Strehl loss is 10\%, which cannot be ignored.
Although focus changes resulting from tracking across the sky can in
principle be calculated, those resulting from changes in the intrinsic
height of the layer -- which can exceed 1--2\,km on longer timescales --
cannot, and must instead be measured.
At least three methods have been proposed to do this, namely direct
imaging of the sodium plume; 
LIDAR (LIght Detection and Ranging); 
and averaging the LGS WFS focus error over 30--60\,s.

The disadvantages of direct imaging of the plume are that an additional
telescope is needed; height resolution depends on both seeing and
distance from the main telescope; the pointing model is complex; and
small errors can result in large errors in height determination (e.g.,
Michaille et al. \cite{mic00}; and section~\ref{Testimaging} of our
paper).
For example, if the monitoring telescope is 200\,m from the main
telescope, the LGS (at zenith) will appear as a plume
$\sim40$\arcsec\ long.
For typical seeing convolved with the intrinsic LGS size, the
resolution of the profile will be limited to $\sim$300\,m. 
Additionally, an error in the pointing of only 1\arcsec\ results in a
height error of 200\,m.
The key advantage is that it can be performed independently of
observations at the main telescope. 

The principal advantage of LIDAR is the accurate height measurements,
and their insensitivity to the ambient conditions.
The disadvantage, however, is the extremely low duty cycle of the laser. 
Since the time between emitting a pulse and detecting the return flux
is of order 1\,ms, the pulse rate of the laser must be slower than
this, and in fact, typical rates are much less than 100\,Hz.
This rules out the use of classical LIDAR during closed loop adaptive
optics, unless a second laser is used.

A third option is to use the focus term from the WFS averaged over
30--60\,s intervals.
This should be zero, and any residual can be attributed to a change in
height of the LGS.
This would be an efficient on-line solution, but there
could be problems with the initial focus setting of the WFS,
particularly with curvature sensors.
Additionally, the use of active optics during AO operation could
induce further errors.
Simulations for the case of the VLT (Bonnet \cite{bon01}) suggest that
the residual rms error should be only 35\,nm.
However, this has been neither tested nor confirmed.

We have developed a method of modulating the beam of a continuous wave (cw)
  laser so
that the duty cycle is 50\% and only relatively small losses occur in
transmission through the extra optics.
The resulting average power remains high enough that the same
laser can in principle be used simultaneously for both a LIDAR
measurement and by an adaptive optics system. 
This is an important consideration since it
is during observations with a LGS-AO system that these measurements are
needed. Additionally, it is worth remarking that by making use of 
 the small amount of light transmitted by AO system mirrors, 
 which is usually unused, both LIDAR detection and the AO system can 
 operate together without
  degrading AO performance; the design proposed for the VLT
  makes use of the 2\% of light lost through one of the 
 mirrors in the AO system.
An experiment employing this technique was conducted with the 
ALFA (Adaptive optics with a Laser  For  Astronomy) AO-LGS system
at the Calar Alto Observatory in Spain during 1999.
The ALFA project was a joint development between the
Max-Planck-Institutes for Astronomy (MPIA, in Heidelberg) and
Extraterrestrial Physics (MPE, in Garching).
It used a cw laser to generate a laser guide star, as this is very
efficient in terms of return flux per Watt of launched laser power.
A Shack-Hartmann WFS, which could be adjusted to the ambient seeing
conditions by switching lenslet arrays, was able to measure the
wavefront from either a natural or laser guide star.

We describe the principle of our sodium profiling method in 
Section~\ref{exp_proc}, followed in Section~\ref{Example} by a
description of the observations, experimental set-up, and results,
including a comparison of LIDAR results with direct imaging.
Looking forward to the fast-approaching era of laser guide star
operation at the VLT, we show in Section~\ref{expected_performanceVLT}
the expected performance of the sodium layer profiler 
(Butler et al. \cite{but02})
to be installed at the Laser Guide Star Facility there.
Finally,  conclusions are drawn in Section~\ref{conclusion_end}.

%__________________________________________________________________

\begin{table*}
\begin{tabular}{ccllll}
\hline 
Minimum N(Na)$^a$ & $<$N(Na)$>$$^a$ & Latitude & Method & Laser & Reference \\
(10$^{-13}$\,m$^{-2}$) & (10$^{-13}$\,m$^{-2}$) \\
%\\
\hline  \\
$\sim$2 & $\sim$3   & 44$^\circ$N$^1$ & LIDAR   & pulsed dye & 
%\hline \\
Megie et al. (\cite{meg77})\\

1.7     & 3.9$^b$   & 41$^\circ$N$^2$ & LIDAR   & pulsed dye & 
She et al. (\cite{she00})\\

   2.3   &    4.3      & 40$^\circ$N$^3$ & LIDAR   & pulsed dye & 
Papen \& Gardner (\cite{pap96})\\

        &           & 37$^\circ$N$^4$ & LIDAR   & cw dye     & 
this paper\\  

$\sim$2 & 3.7       & 32$^\circ$N$^5$ & spectroscopy$^c$ & cw dye     & 
Ge et al. (\cite{ge98})\\

2.6     & 3.6       & 29$^\circ$N$^6$ & imaging & cw dye     & 
Michaille et al. (\cite{mic01})\\

        &           & 20$^\circ$N$^7$ & LIDAR   & pulsed dye     & 
Gardner et al. (\cite{gardner91}), Kwon et al. (\cite{kwo88})\\

        &           & 18$^\circ$N$^8$ & LIDAR   & pulsed dye & 
Hecht et al. (\cite{hec93}), Collins et al. (\cite{col02})\\

$\sim2$ & $\sim4$   & 23$^\circ$S$^9$ & imaging & pulsed dye & 
Simonich et al. (\cite{sim79}), Clemesha et al. (\cite{cle82})\\

2       & 3.6       & 30$^\circ$S$^{10}$ & imaging & cw dye     & 
D'Orgeville et al. (\cite{org02})\\
\hline \\
\\
\end{tabular}

$^a$ Minimum and mean sodium column densities, measured at zenith.

$^b$ Average of reported maximum and minimum (6$\times$\,10$^{-13}$\,m$^{-2}$ and 1.7$\times$\,10$^{-13}$\,m$^{-2}$ respectively)

$^c$ Spectroscopy of bright stars, to look at telluric sodium
absorption, was used to determine Na abundance; this was compared to
the laser guide star flux.

$^1$ Observatoire de Haute Provence; 
$^2$ Fort Collins;
$^3$ Urbana;
$^4$ Calar Alto Observatory;
$^5$ Kitt Peak National Observatory;
$^6$ Observatorio de Roque de los Muchachos;
$^7$ Hawai`i;
$^8$ Arecibo Observatory;
$^9$ S\~an Jos\'e dos Campos;
$^{10}$ Cerro Tololo Inter-American Observatory.

\caption{List (non-exhaustive) of sites where mesospheric sodium layer
structure and/or 
 abundance has been measured, and the type of laser used for the measurements}
\label{min_abundances}

\end{table*}

\section{Principle of sodium profiling by LIDAR}
\label{exp_proc}

When using sodium fluorescence to measure the sodium density profile, 
it is important that we know the relation between the two quantities,
which can depend on both the laser intensity and the sodium column density.
If the laser intensity in the mesosphere is higher than 
6\,W\,m$^{-2}$\,MHz$^{-1}$ then saturation losses will reach 50\%. 
Since this is caused by the finite decay time of excited sodium atoms, 
it depends only on the laser intensity.
For a 4\,W laser like ALFA (cw, 10\,MHz bandwidth) modulated with 50\%
efficiency so that the maximum power launched during a LIDAR
measurement is only 2\,W, a 1.5\arcsec\ spot would not saturate the
sodium.
The same is true for PARSEC (Rabien et al. \cite{rab02}; Davies et
 al. \cite{dav02}), the laser being built for the VLT. During a
LIDAR measurement it will launch up to 8\,W cw, with 4 single frequency
lines within a 500\,MHz envelope.
For such cases, we can be sure that the fluorescence and density
profiles of the mesospheric sodium have a direct relation.
Additionally, the sodium column density is low enough  
($\sim3\times10^{13}$\,m$^{-2}$) that very little of the laser power
is absorbed (only a few percent at zenith, and still less than 10\% at a
zenith distance of 60$^\circ$), so the laser intensity does not change much
through the layer.
%, unlike the sodium fluorescence.

Classical LIDAR has been used to study  the sodium layer for many years 
(see Table~\ref{min_abundances} and references in 
Clemesha \cite{cle95}, Ageorges et al. \cite{age00}, Sica et
al. \cite{sic02}, and Collins et al. \cite{col02}).
It consists of sending short laser pulses of the order of
0.1--1.0\,$\mu$s at a rate of a few Hz and collecting the returned
photons in time bins of similar width to the outgoing laser pulses. 
By combining the returned photons from many laser `shots', 
the backscattering profile  of  the atmosphere   is obtained; 
the signal-to-noise depends on the number of shots.
Although the energy in each pulse is up to a few $\times10$\, mJ,
saturation is avoided by illuminating a large area (20--200\arcsec) of
the sodium layer; such intrinsically pulsed sodium  lasers 
 are created by intra-cavity modulation  (called Q-switching) -
 energy is stored up, and transmitted in a pulse. 
Additionally, pulse separations greater than 700\,$\mu$s at zenith
(and 1400\,$\mu$s at a zenith distance of 60$^\circ$) are required to
avoid confusion over the location of the photon emission by overlapping
of the return flux (both Rayleigh/Mie scattered from $\sim$\,15-25\,km and
fluorescent from $\sim$\,80-100\,km) from successive pulses.
Such lasers cannot be used for simultaneous adaptive optics
correction, which needs a compact laser spot on the sky and operates at frame
rates up to $\sim1$\,kHz.

 The alternative would be intra- or extra-cavity modulation of 
 an existing cw laser. However, for adaptive optics correction and 
 sodium layer profiling in parallel, the low duty cycle
  of the classical LIDAR approach  again  provides  
 too few return sodium layer photons,  which would anyway be swamped by 
 background noise from sky photons, dark noise, and stray photons. 
  Importantly though, for the VLT sodium layer profiling system, 
 the beam on the sky is focused into a 1.5\arcsec\ spot 
 and observed through a small field aperture which cuts down the 
 sky photon flux.

\begin{figure}
\centerline{\includegraphics[height=9.0cm,width=8.9cm,angle=0]{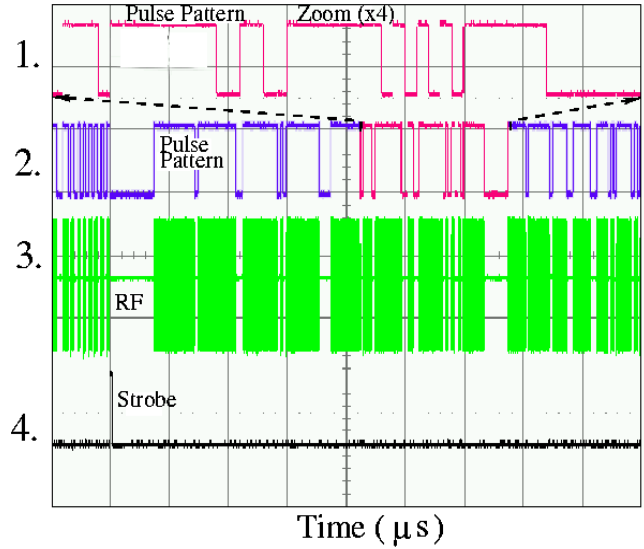}}
\caption{Example section of a LIDAR electrical sequence generated by a 
 pulse pattern generator and the associated radio frequency (RF) signal
 from the acousto-optical modulator driver.
  Channel 2:
 a 200$\mu$s snapshot of random 1$\mu$s TTL-pulses in  `non-return to zero'
  format (at 2.0\,V peak-to-valley).
 Channel 1: 4 $\times$ zoom of channel 2. 
 Channel 3: 200$\mu$s snapshot of the  40\,MHz RF signal. 
 Channel 4: Strobe signal indicating the start of a sequence.
 }
\label{fig:pulse_pattern}
\end{figure}

 For laser modulation at high duty cycles, optical modulation is perhaps
 more  convenient than the mechanical solution, a rotating chopper,
  as it allows  a range of laser pulse widths and  spacings. In 
  acousto-optical modulation, for example,   
 applying a radio-frequency  signal   to the modulator
 causes most of the laser light to be 
 diffracted out of the straight-through beam  into
  different directions. When the signal stops, the beam reverts
 to being  straight-through.  By using an aperture 
  which allows only one of the beam directions to be transmitted
 (i.e. a laser beam is either all 
 transmitted or all blocked), one can create digital On/Off pulses. 
 Because a sequence of two or more  high states 
  is  continually high, the pattern has a 
   so-called `non-return to zero' format (see Fig.~\ref{fig:pulse_pattern}).

To maximize the sodium layer return flux, we have adopted this 
 technique in order to launch a pseudo-random string of laser pulses
 with a `non-return to zero' format, providing a laser 
 duty cycle of 50\%. The sequence 
can be repeated many times without pausing (as long as the
time taken to complete one sequence is longer than the maximum round-trip time
for a photon) and the return flux is recorded as a function of time
in a series of consecutive measurement channels.
Each time the sequence restarts, recording begins
again in the first channel, adding to the counts already there.

If $S_0$ is the out-going laser pulse sequence then the returned stream of 
photons results from  the convolution of 
$S_0$ with the sodium profile $N \otimes S_0$.
The intrinsic sodium abundance profile, $N$,  can be recovered 
from the data by cross-correlating 
it with the original pulse sequence if the auto-correlation of the 
sequence is very close to being a delta-function.
Thus we find

\begin{equation}
S_0 \otimes ( S_0 \otimes N ) \ = \ N\,.
\end{equation}

For the ALFA experiment, we used a variation of this by over-sampling,
so although the pulses were 1\,$\mu$s long we collected the returned
photons in 0.25\,$\mu$s time bins.
Now to recover the sodium profile we consider the emitted laser pulse 
sequence as a sequence of impulses, $S_1$, four times 
longer in which each digit of $S_0$ is padded with zeros.
If the profile of a pulse from the laser is denoted by $L$ then we can 
consider the emitted 
sequence as $S_1 \otimes L$, and the returned flux is 
$S_1 \otimes L \otimes N$.
We can calculate the following cross-correlation 
\begin{equation}
S_1 \otimes ( S_1 \otimes L \otimes N ) \ = \ L \otimes N\,, 
\end{equation}

which gives the convolution of the sodium profile with the pulse 
profile.
For the VLT LIDAR measurements, we will use Nyquist sampling and
record the return flux in time bins equal to half of the pulse length;
otherwise the method is similar.

Finally, a correction has to be made to this profile to compensate for the 
height at which each photon was scattered because the telescope mirror 
subtends a smaller solid angle for emission that originates higher in the 
atmosphere. This correction is applied by multiplying the profile
 data by the square of the height.

\begin{figure}
\centerline{\includegraphics[width=8.4cm]{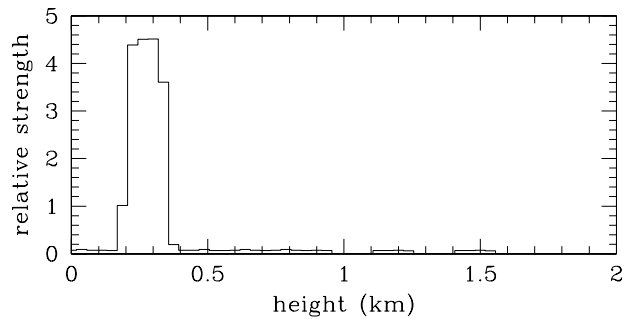}}
\caption{The profile of the 1\,$\mu$s laser pulse, measured by
scattering the beam from the telescope dome. 
The profile is close to being square-wave and has a FWHM of 150\,m, as
expected.
The height offset of 270\,m is due to a delay of 1.8\,$\mu$s between
the pulse generator, modulator, and detector.
It has been subtracted from all subsequent height measurements.}
\label{fig:pulse_profile}
\end{figure}

The shape of the pulse profile, and any height offset which might arise due 
to timing delays, can be found by carrying out the procedure with the 
telescope dome closed: this provides a single scattering layer at almost 
zero distance.
As an example, the cross-correlated dome data from the ALFA LIDAR
observations in Fig.~\ref{fig:pulse_profile} shows that
there is a height offset of 270\,m. 
This is due to a  timing delay in the system of 
1.8\,$\mu$s between the pulse generator, modulator, and detector, 
 and has simply been subtracted from all other derived heights.
It can be seen that the laser pulse has a 
form close to a square-wave with a measured width of 
150\,m, which is what we would expect for a 1\,$\mu$s pulse with a
rise time of less than 100\,ns.
For high signal-to-noise data, this profile can be used to deconvolve the 
sodium layer profile to yield a height resolution better than 100\,m.

%----------------------------------------------------------

\section{Sodium Layer profiling by LIDAR at Calar Alto Observatory}
\label{Example}

\subsection{Experimental set-up}
\label{Expt_set_up}

\subsubsection{The laser} 

The ALFA laser system (Rabien et al. \cite{rab00}), situated 
in the Coud\'e laboratory of the 3.5\,m telescope at Calar Alto
Observatory in Spain, has been used to 
provide an artificial reference star in the mesospheric sodium layer
for adaptive optics correction.
It could generate a LGS with a magnitude as bright as V$\sim$10 in
good atmospheric transparency.  
The laser system consisted of a ring dye laser (Coherent 899-21,
with some modifications) pumped by a 28\,W cw Argon ion laser
(Coherent Innova 400).
The output power could exceed 5\,W in single mode with a 10\,MHz
bandwidth.
The frequency was tuned to the sodium D$_2$ line  at 589\,nm. 
The laser beam was circularly polarized, pre-expanded, 
and sent to the 50\,cm diameter laser launch telescope, 2.9\,m from
the science telescope optical axis, via
a remotely controllable series of relay 
mirrors (Ott et al. \cite{ott00}). 

\begin{figure}
\centerline{\includegraphics[width=8.4cm]{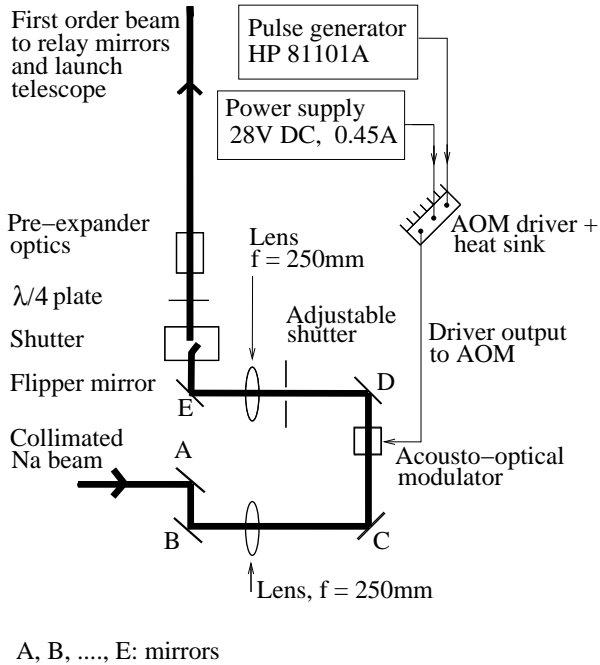}}
\caption{Schematic of the set-up in the laboratory used to modulate
the laser beam. The thick line shows the beam path.}
\label{fig:laser1}  
\end{figure}

\subsubsection{Laser modulation}\label{laser_modulation}

An ISOMET 1201E-1 acousto-optical modulator (AOM), driven 
by a pulse generator (HP~81101A) provided the optical beam modulation.
As described in Section~\ref{exp_proc},  close to square wave amplitude 
modulation of the laser beam at MHz rates 
was desired to achieve the required height resolution and a maximum
returned flux.  
It is the pulse rise (and fall) time that governs the squareness of a
pulse and in the AOM the rise time is dependent on beam diameter.  
To achieve faster rise times it is  necessary to
focus the laser beam, so that in the crystal the acoustic transit time
across the beam is reduced.
The laser modulation set-up is shown in Fig.~\ref{fig:laser1}.
In order to achieve a pulse rise time below 100\,ns,
the laser beam was focused by a lens placed about 1\,cm in front of
the modulator.
The resulting power density of less than 300\,W\,mm$^{-2}$ was well
below the limit of 800\,W\,mm$^{-2}$ for the modulator.
A second lens approximately compensated for the beam quality degradation
introduced by the modulation optics.
As seen in Fig.~\ref{fig:laser1}, three mirrors 
steered the beam around the additional loop and
then back onto its nominal path.
An adjustable shutter placed between the lenses 
and after the modulator was used to block all spots except the 
chosen first order spot. 
Some second order light passed through to the beam expander but this
was largely blocked by its beam stop aperture.   
The first order beam power was maximised by following the recommended
procedure: to reduce the driver output power close to
its minimum, then find the optimal (i.e. Bragg) angle of the incident laser
on the crystal by monitoring an optical power meter, and finally set
the driver output power that maximises the first order beam optical
power.
  
A shift in the frequency of the beam arises from modulation of the
beam into several orders.
Within the modulator, the acoustic waves which create the diffraction
pattern that diverts the beam move through the medium, and the
frequency of the diffracted beams is doppler shifted with respect to
that of the incident beam. 
If light is incident at the Bragg angle, the frequency shift
is -70\,MHz at $\lambda = 589$\,nm. 
This is an acceptable change because the width of the D$_2$ resonance
line is $\sim1$\,GHz.

We achieved a relatively low modulation efficiency, with only
about 70\% of the power transmitted in first order.
Including the beam relay transmission, and taking into account the
duty cycle of the pulse sequence, the average laser power at the
launch telescope was $1.1\pm0.1$\,W for the experiment.

\subsubsection{Detector set-up}
\label{detector}

The accuracy of LIDAR data is limited by the timing resolution of the detector.
For this reason CCD detectors are not suitable, and  
photo-multiplier tubes (PMTs) are more commonly used. 
However, PMTs have a low quantum efficiency (QE $\sim10$\%)
and instead we have used an  
actively quenched silicon Avalanche PhotoDiode (APD) from Perkin Elmer
 with a  QE greater than 70\%. 
Actively quenched APDs  minimize the 
dead time after each detected photon to $\sim20$\,ns.
Statistical simulations show that we only expect a maximum of 
 one to two photons from the laser beacon to be detected every micro-second. 
Therefore, the dead time does not affect the measurements. 
The APDs have no  readout noise and the dark counts are
$<1000$\,s$^{-1}$; the dominant source 
of noise is background counts, including sky counts and stray light
around the sodium line transmission filter in the optical path in
ALFA.
In addition to the usual dust and light shields surrounding 
the  bench, heavy black cloth was attached to ALFA to block
stray external optical light from reaching the APD.
However, much more serious was the presence of infrared LEDs
inside the ALFA bench, on motor encoders for example. 
Although every effort was made to remove these the 
overall background count rate was $\sim7000$--8000\,s$^{-1}$. 

\begin{figure}
\centerline{\includegraphics[angle=-90,width=8.4cm]{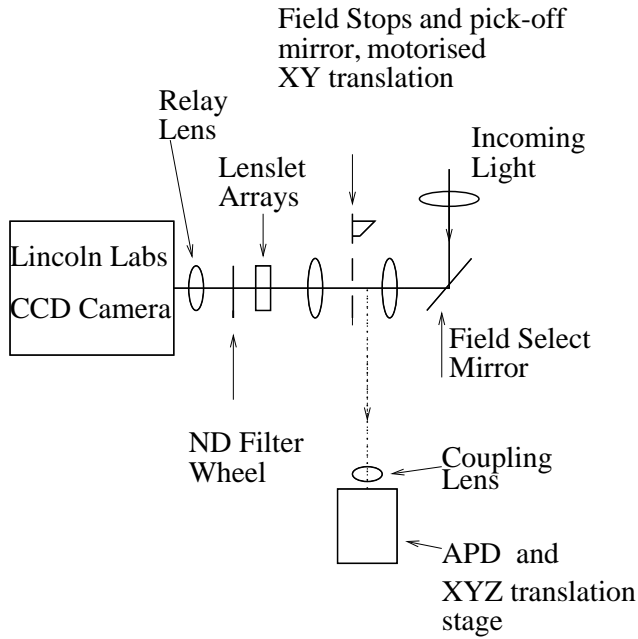}}
\caption{Schematic of the set-up of the APD LIDAR detector close to the
Shack-Hartmann wavefront sensor in ALFA.
No neutral density (ND) filter was used during LGS LIDAR
observations.}
\label{fig:shs}
\end{figure}

In mounting the detector on the telescope we only had access to the F/25 
focus close to the wavefront sensor, at which the plate scale is
0.42\,mm\,arcsec$^{-1}$. 
A 12$\times$ magnification of the image of the APD onto 
the F/25 focus using a lens was required to obtain a 6\arcsec\ FoV (to
match the typical size of $<3$\arcsec\ for the LGS).
Given a 200\,$\mu$m diameter  APD active area, the  plate scale  
 at the APD  was therefore $\sim0.03$\,mm\,arcsec$^{-1}$. 
Alignment of the optical system was performed using a white fibre source
placed at the F/10 telescope focus to simulate the laser guide star.
As seen in Fig.~\ref{fig:shs}, the beam could be directed either to the WFS 
or the APD by a mirror on a remotely movable  translation stage. 
Note that this experimental set-up differs significantly to that
proposed for the VLT, in which the 2\% of light leaking through one of
the mirrors in the AO system will be used for the LIDAR measurements.

\subsubsection{Data collection}

The APD module outputs a TTL pulse every time a photon is received. A similar 
TTL sync signal is produced from the signal generator at the start of every 
16k bit pseudo-random sequence (the sequence provided with the pulse
generator was in fact a 4k bit sequence repeated 4 times). 
These two signals are received by a 
 multichannel scaler (MCS) where timing information is recorded.
The MCS card (Fastcomtec MCD-2) consists of 128k channels, each of which 
corresponds to a 250\,ns interval. 
When the sync pulse is received the MCS starts 
at the first channel, and increments its value every time it
receives a count from the APD during the subsequent 250\,ns. 
After this time it moves to the next channel and 
again counts the number of pulses coming from the APD. This procedure 
continues until the next time a sync pulse is received, i.e. when the MCS starts from the first channel again.
In this experiment we over-sampled, with a time interval per channel
one quarter of a pulse length.

The MCS has a dead time between channels of 0.5\,ns. 
For 3.5-m and larger telescopes this is  not 
significant for the APD channel as it has the same effect as reducing 
the collecting area of the telescope by less than 0.2\%. 
Its effect can occasionally be 
seen in the sync channel, and this is why we have twice as many channels as 
should ever be needed:
occasionally the MCS will miss the sync pulse and the 
collected data will `overflow' into these other channels. 
Although this data is not used, this precaution prevents the rest from being corrupted.

\subsection{Laser star acquisition}

A 2D detector is required for time efficient laser spot acquisition. 
Prior to the observations, the APD was aligned to the white light
reference beam by adjusting the APD stage  to maximize the count rate
on the APD event counter. 
At the beginning of the observations, the laser was 
pointed to its  nominal position on the WFS. 
One of the white light fibre spots on the WFS
 was  then marked and  the corresponding sodium laser WFS spot
 was moved to this mark. 
Following this, a pick-off mirror was moved to
direct the LGS beam to the APD (Fig.~\ref{fig:shs}),
and integration was started. 

% ------------------------------------

\subsection{Sodium profiles from LIDAR}
\label{results_CA_data}

\begin{figure*}
\centerline{\includegraphics[width=8.4cm]{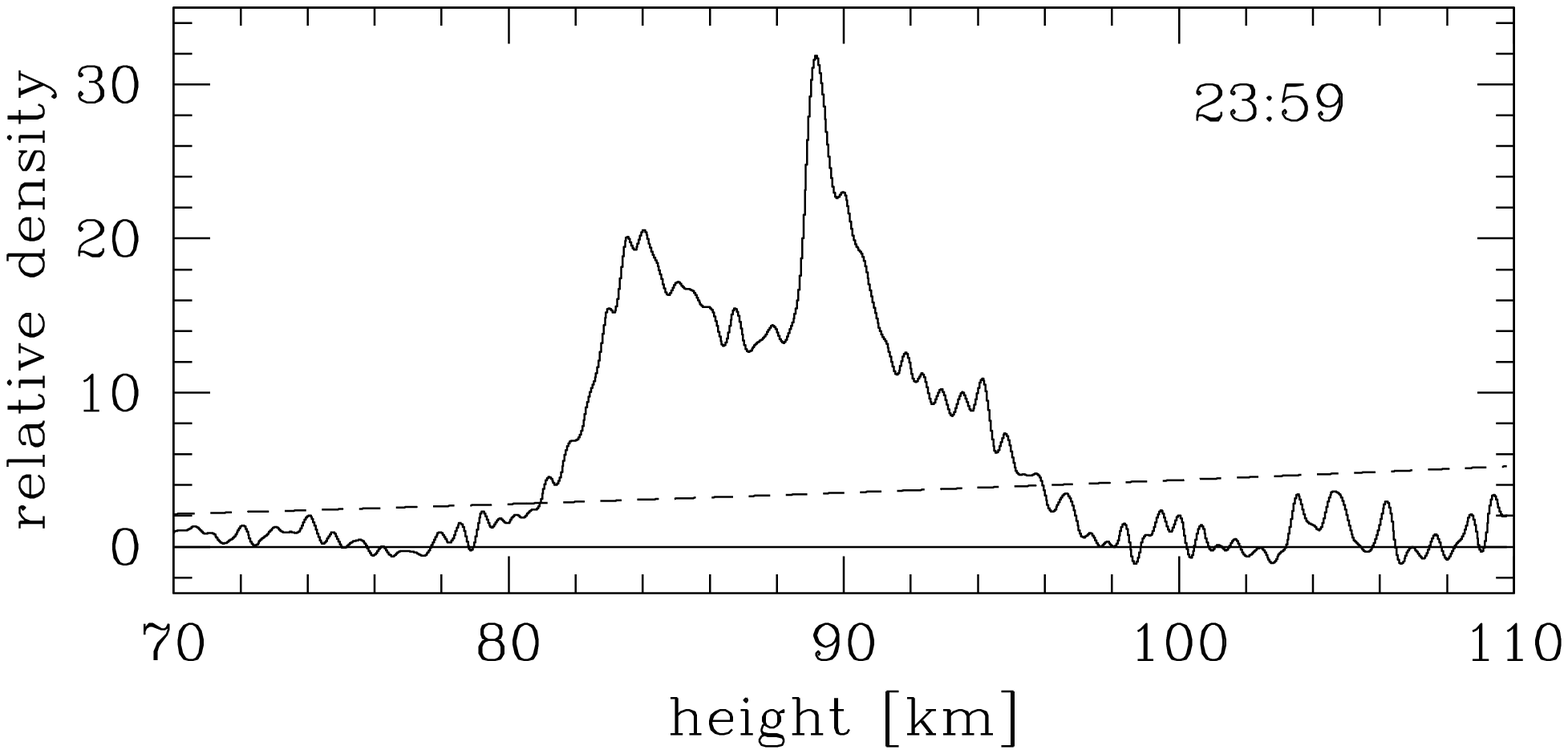}}
\centerline{\includegraphics[width=8.4cm]{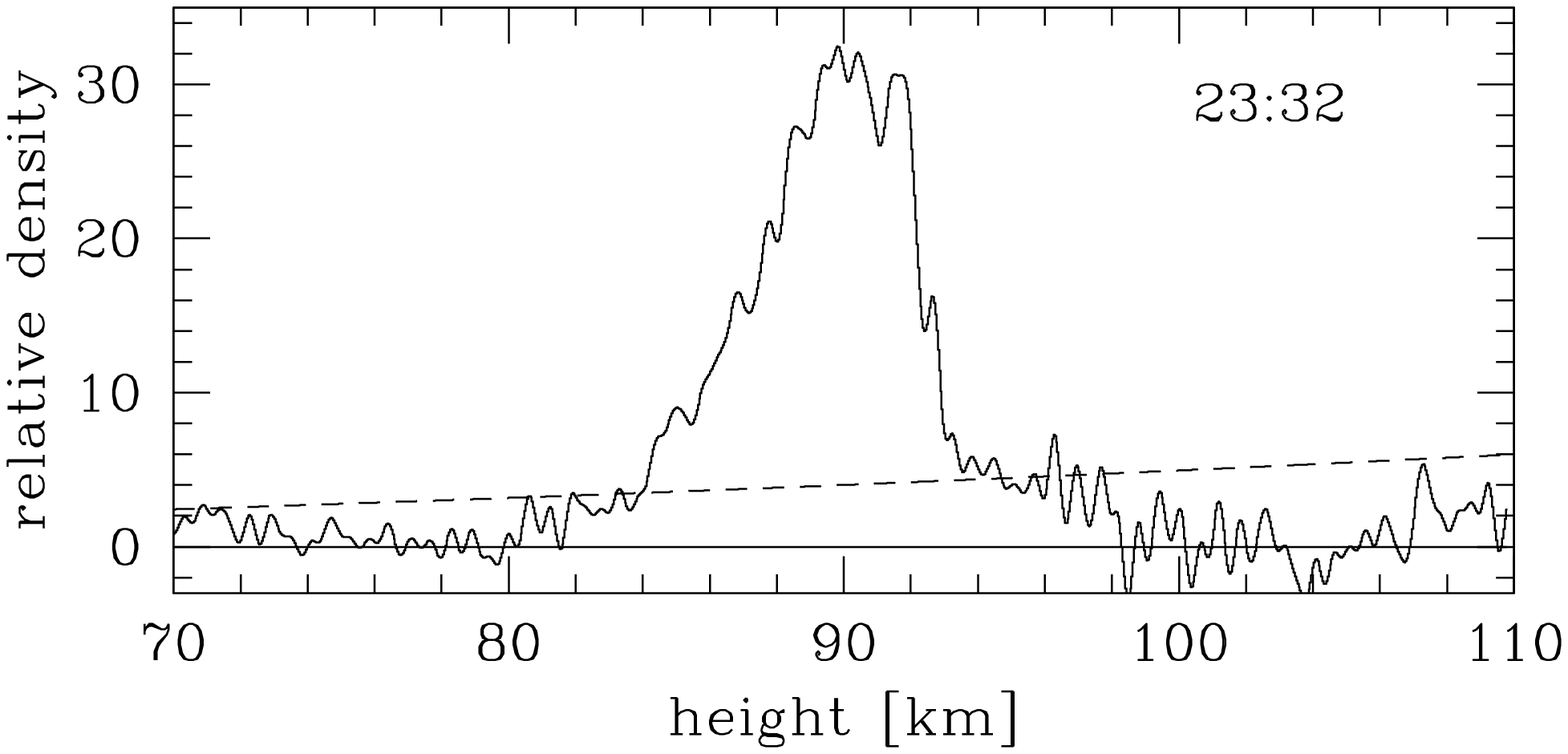}}
\caption{Examples of profiles of the sodium layer taken on two
consecutive nights: Oct 17 (upper) and Oct 18 (lower).
The UT start time of each 30\,sec integration is shown.
The 3$\sigma$ noise level (dependent on height) is shown as the dashed line;
details of how it is determined are given in the text.
Heights are given in km above Calar Alto Observatory which is itself at
2.2\,km. 
The difference in profile between the two nights is substantial.
\label{fig:na_profiles}}
\end{figure*}

The first measurements using this LIDAR technique with ALFA were
obtained in 1999 on October 17 \& 18 under 
extremely poor weather conditions.
On the first night the seeing was 4\arcsec\ or more. 
Even though the laser had been launched with a 15\,cm beam, in such bad seeing 
this is much more than a few times the Fried parameter, $r_0$,  
 and so the size of the LGS is affected.
The final size of the LGS as seen from the ground was at least 6\arcsec. 
As a result, significant light was lost from the 6\arcsec\ field of
view (FoV) of the APD.
On the second night there were thin cloud layers at 6\,km and 9\,km 
(observed by the LIDAR system from the light they scattered).
LIDAR integrations were UT time-tagged for comparison  with sodium
layer profiles derived from direct imaging at a second telescope (see 
 Section~\ref{Testimaging}). 
Simultaneous observations of the LGS from this second telescope showed that 
the observed flux was reduced by a factor of thirty by these cloud layers.
The only effect this has on LIDAR is to reduce the signal-to-noise and 
invalidate absolute flux calibration; 
the height and profile information (which arise solely from time-tagged 
data) are unaffected, and thus the first results, shown in 
Fig.~\ref{fig:na_profiles} are very encouraging.
These data have been convolved with a low-pass digital filter to give a 
smoothing of 500\,m, and only the range encompassing the mesosphere plotted.
Additionally, the total flux detected has been normalised for each night, 
a process which is not normally necessary but which we use in this case 
because of the significant flux variations due to the clouds.

The Rayleigh cone, which is bright at altitudes less than $\sim$20\,km was 
not detected. This is because we had a  FoV  of 6\arcsec\ and  launched the
laser 2.9\,m off-axis from the main telescope,  which means we 
 could only observe heights in the range 90$\pm$30\,km.
The reason we were able to observe cloud layers below 10\,km is simply 
because there was so much scattered light from the defocused image of the 
pupil on these layers, that we could detect it about 1\,arcmin away.

The noise is strongly dependent on height, because of the necessary
scaling described in Section~\ref{exp_proc}.
It is determined from the data before the scaling is applied
(i.e. while the noise is independent of height) and then scaled in the
same way as the data -- that is, multiplied by the square of the height.
The result is that the noise at 100\,km is 25\% more than (and 
at 80\,km 20\% less than) that at 90\,km.
The blank region of the profile used to make the estimate is at 10--60\,km,
which is large enough that both the photon and correlation
noise are implicit in the estimate.

\subsubsection{Results and discussion}
\label{sect_results}

\begin{figure*}
\centerline{\includegraphics[height=7cm]{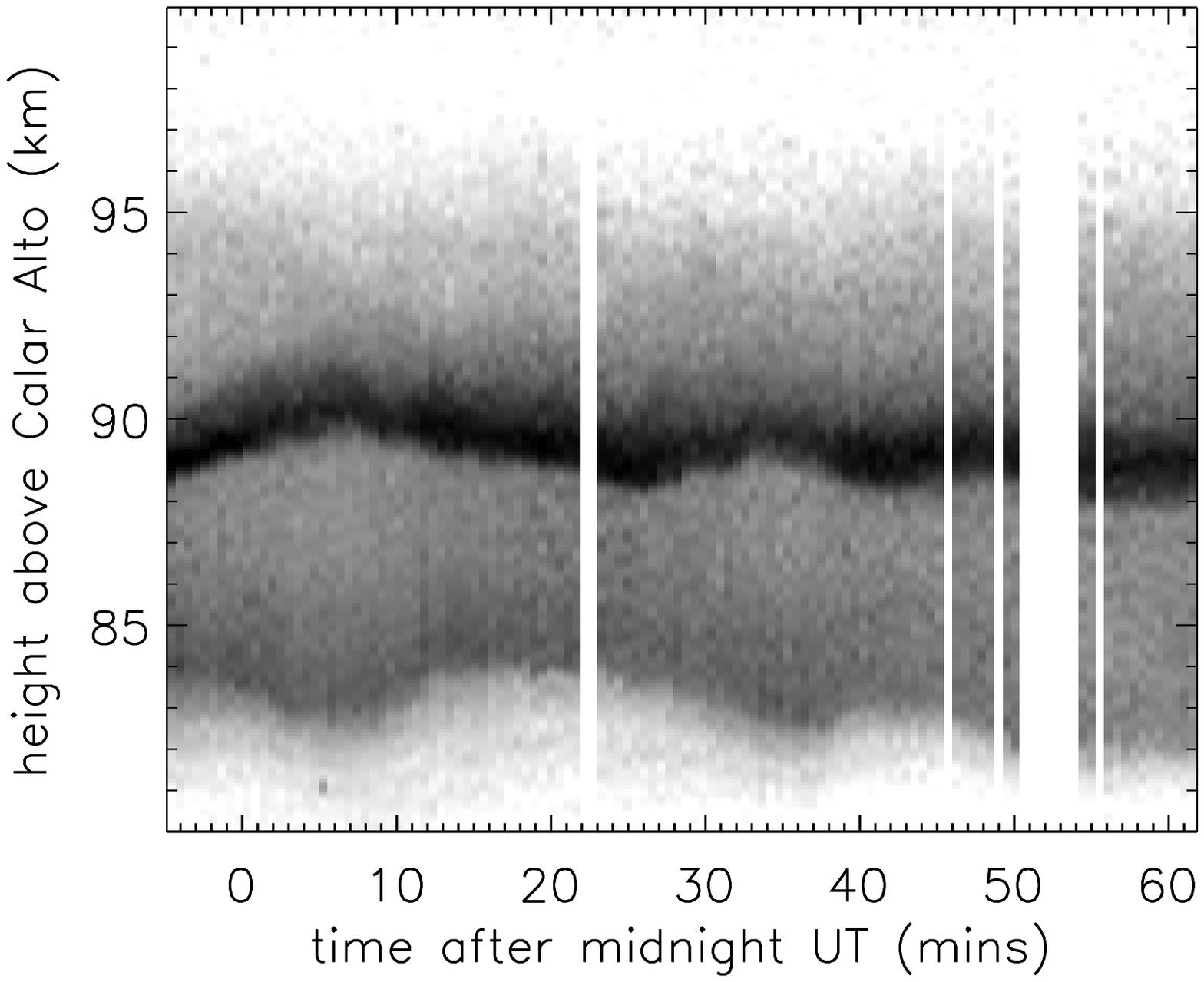}\hspace{5mm}\includegraphics[height=7cm]{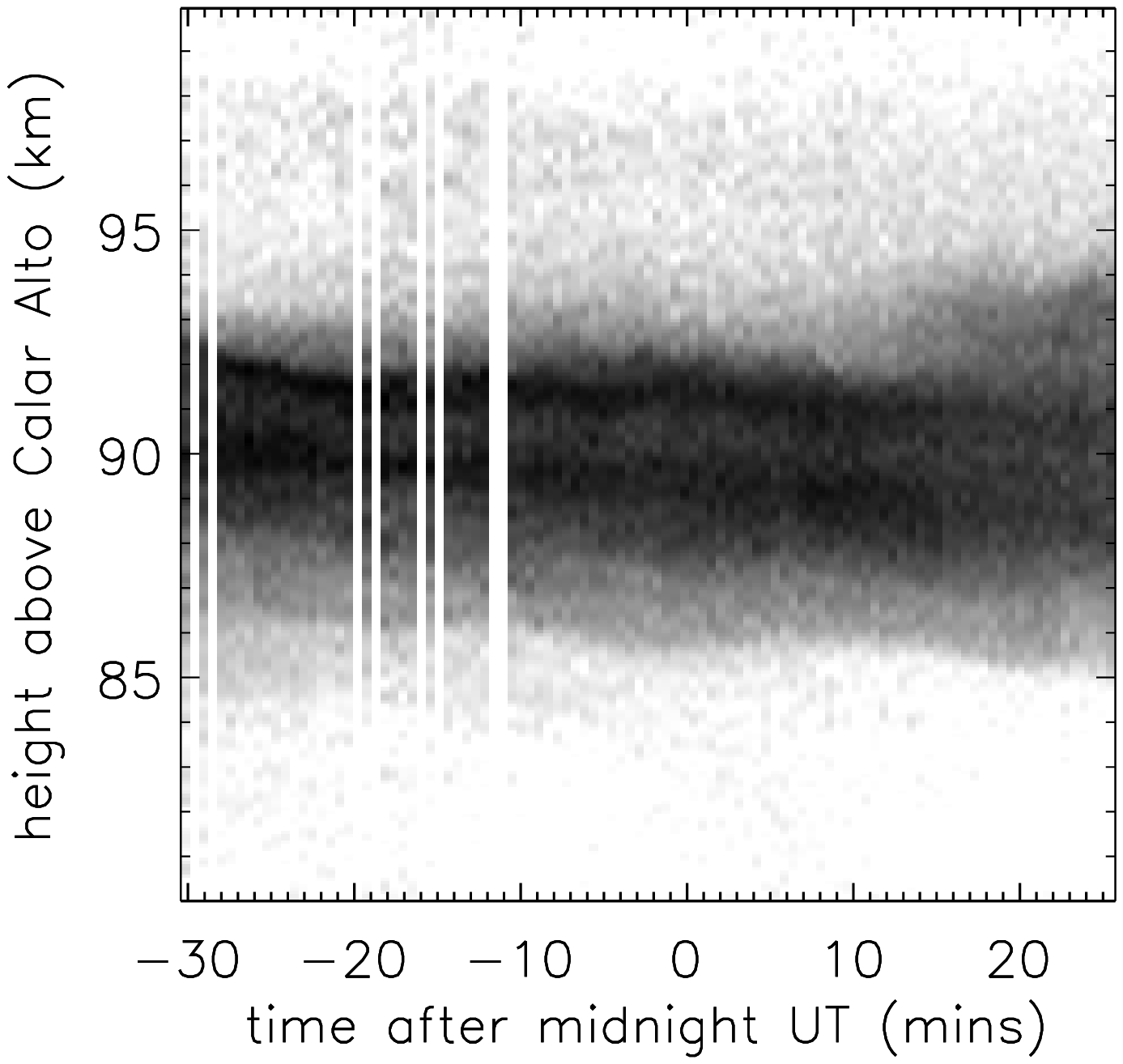}}
\caption{Density profiles of the sodium layer on Oct 17 (left) and 
Oct 18 (right) 1999, taken every 30\,s over a period of 1 hour each night.
Darker regions denote a higher atomic density.
Blank columns indicate where no data was collected due to clouds.}
\label{fig:full_na_oct18_19}
\end{figure*}

\begin{figure*}
\centerline{\includegraphics[width=8cm]{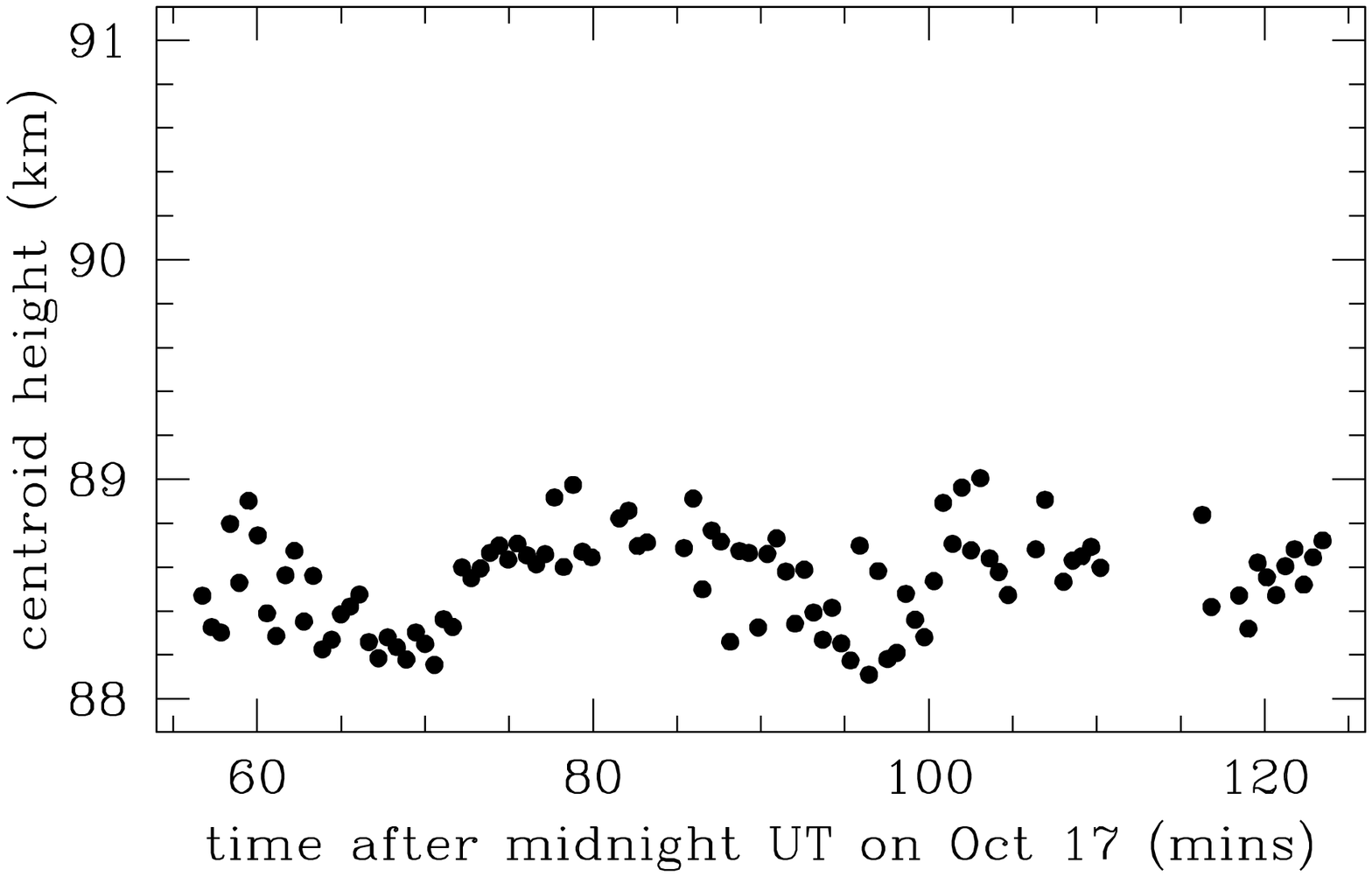}\hspace{5mm}\includegraphics[width=8cm]{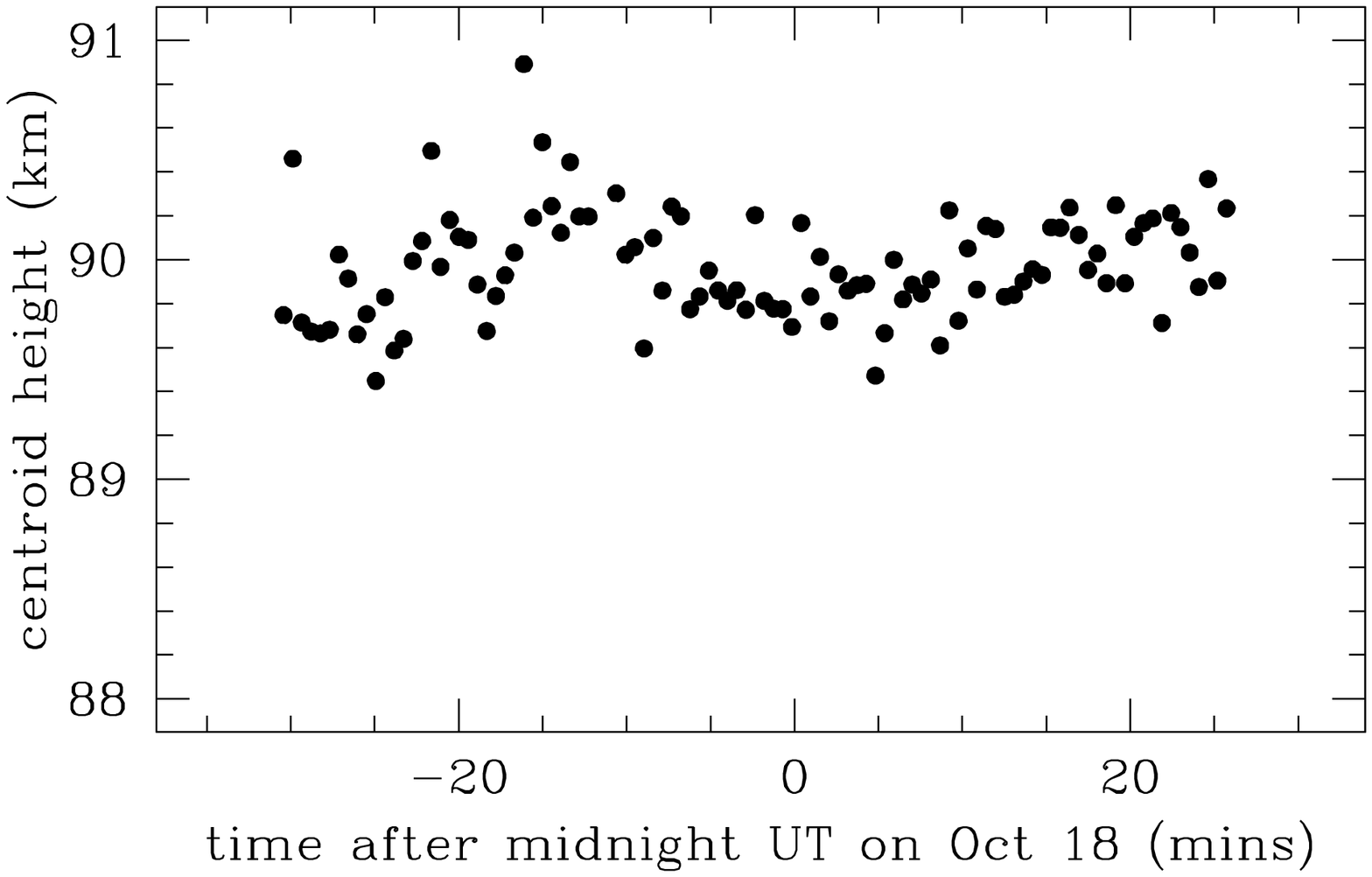}}
\caption{Plots of centroid height of the sodium layer over time on
Oct 17 (left) and Oct 18 (right) 1999.  
Heights are given in km above Calar Alto Observatory which is itself at
2.2\,km.
There was a 1.5\,km difference in height between the two nights, and
further smaller variations during each night.}
\label{centroid-na}  
\end{figure*}

It can be seen from Fig.~\ref{fig:full_na_oct18_19} that using this
method we are able to track changes in the sodium layer structure on
timescales as short as a few minutes.
The two important quantities for adaptive optics are the centroid
height and the width over which the layer is spread.

The intensity-weighted centroid height of the atoms shown in
Fig.~\ref{centroid-na}, suggests that general trend is for variations
of order 500\,m on timescales of 10\,minutes.
It is not clear from these data whether the changes between individual
points are real effects or a result of the noise in the data (due to
the poor conditions under which the experiment was performed).
 Comparison with the variations in centroid height derived 
 from direct imaging by
O'Sullivan et al. (\cite{sul00}) at the same observatory during 1998
 would suggest that these could indeed be real variations.

Additionally, there is no correlation between
properties of the layer between the two nights.
On Oct 18, the layer (excluding the sporadic) had a distribution which
appears almost bimodal, with a centroid height of 92\,km above mean
sea level;
on Oct 17, the centroid height was only 91\,km, and there was a very
prominent narrow peak.
Because it had a FWHM of only a few hundred metres and moved
independently from the rest of the layer, we identify this as a sporadic layer.
Within the sporadic layer, the sodium density is about twice that of the
underlying profile, similar to the  factor of two to three observed 
during the O'Sullivan et al. (\cite{sul00}) direct imaging observations.

The FWHM of the layer also shows significant changes.
On Oct 17, the layer had a FWHM of around 12--13\,km; on Oct 18 this
had reduced to only 5--7\,km. 
This has important consequences on the elongation of the LGS
spot as seen from different portions of the science telescope primary
mirror, and hence on the performance of an AO system correcting on the
LGS.
For a laser projected from the side of a 10-m primary mirror, as
occurs at the Keck telescope, this can result in elongations in the
range 1.5--3.1\arcsec\ for apertures located on the far side of the
pupil.
On nights when the layer has a large FWHM, the performance of an AO
system correcting on the elongated LGS will be compromised;
not only due to the shape of the spot, but also because it may be
truncated by the limited aperture of the wavefront sensor, and also
due to the reduced signal to noise in the measurement of the spot
positions. 
At the VLT and Gemini, it  the laser will be projected from
behind the secondary mirror.
This reduces the maximum elongation to the range 0.6--1.3\arcsec\ for
the cases above.
When convolved with both the seeing and the intrinsic size of the LGS
(combined, about 1.2\arcsec), this has only a small effect on the
shape of the LGS.

%------------------------------------------------------------

\subsection{Comparison to Direct Imaging}
\label{Testimaging}

In order to validate the profiles obtained using the LIDAR method, we
simultaneously observed the LGS plume from another telescope.
For this we used the 2.2-m telescope, separated from 
the 3.5-m by 260\,m.
The detector was a 2048$\times$2048 pixel SITE\#1d
CCD camera with a pixel scale of 0.53\arcsec, installed at its
Cassegrain focus.
A narrow band interference sodium filter was used to minimize the
background.
Images, each a 30\,s exposure, were time-tagged with the UT start time.
The purpose of the observations was multifold. The monitoring of the
sodium spot was meant to detect possible sporadics and determine the
centroid height variation. The hope was to achieve photometric study
of the Rayleigh scattering and sodium spot.
Therefore spectrophotometric standards have been observed
regularly. Unfortunately the weather conditions were not photometric
and the results can therefore only be indicative.
Here we consider only a comparison with the LIDAR data for which, to
ensure it is meaningful, only
images acquired within 30\,sec of LIDAR data are used.

\begin{figure}
\centerline{\includegraphics[angle=-90,width=6cm]{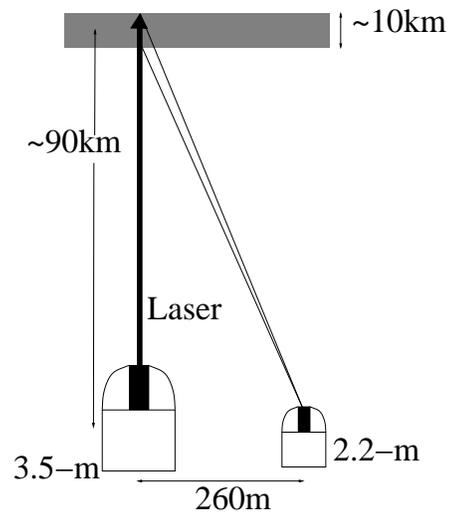}}
\centerline{\includegraphics[height=9.7cm,angle=-90]{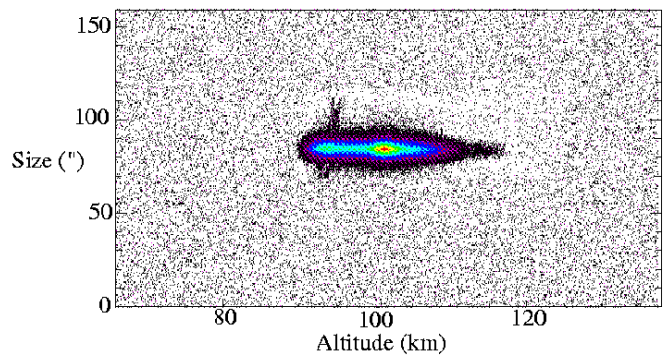}}
\caption{Top: Sketch showing the geometry of the laser guide star
projected from the 3.5-m, and the direct imaging observations from the
2.2-m 260\,m away.
Bottom: Typical image of the ALFA LGS plume as observed from the 2.2-m
telescope. See Section~\ref{Testimaging} for details of the data reduction.}
\label{ima-na}  
\end{figure}

Since no sky frames were acquired, no sky subtraction could be
applied. Only a dark of equivalent integration time was subtracted
before flat-fielding with a dome flat.
Bad pixel correction was also applied.
In a cosmetic final stage, star trails crossing the plume or Rayleigh
beam were removed by thresholding the data.
Fig.~\ref{ima-na} illustrates a typical result.

Due to the geometrical situation of the telescopes and  the fact that the altitude is aligned with the horizontal axis in the images, the altitude of a point can be determined as: 

\begin{equation}\label{eqn1}
{\rm Altitude} = z_0 +  [d* \tan (\alpha - (x_0 - x) * pixscale )] \,, \end{equation}

where $z_0$ is the altitude of the observing site (2.2\,km), $d$ the distance
between the telescopes, $\alpha$ the elevation of the 2.2-m telescope, and
$pixscale$ is the pixel scale of the camera used. 
$x_0$ has been chosen as the middle of the CCD FoV. 
Using this formula, we noticed a difference in
altitude, and altitude range, between the imaged and LIDAR profiles.
To obtain a similar altitude range for both sets of data, the 2.2-m
pointing had to be considered 7.12\,arcmin higher than that actually
indicated at the telescope. 
This can be attributed to the fact that
(a) the 3.5-m (and laser) was pointing to nominal, but not necessarily
actual, zenith;
(b) for observations of the LGS, the 2.2-m pointing was measured as an
angular offset from nominal zenith which again may not necessarily be
actual zenith; and 
(c) there was a repeated offset of nearly 6\,arcmin from the CCD frame
centres when pointing to calibration stars, suggesting a large error
in the pointing model of the telescope.

Once the data was calibrated in altitude, a scaling calibration was
applied:  
the integrated flux of the 2.2-m data between 70 and 110\,km was
forced to equal the integrated flux of the LIDAR profile in the same
altitude range.

\begin{figure*}
\centerline{\includegraphics[width=17cm]{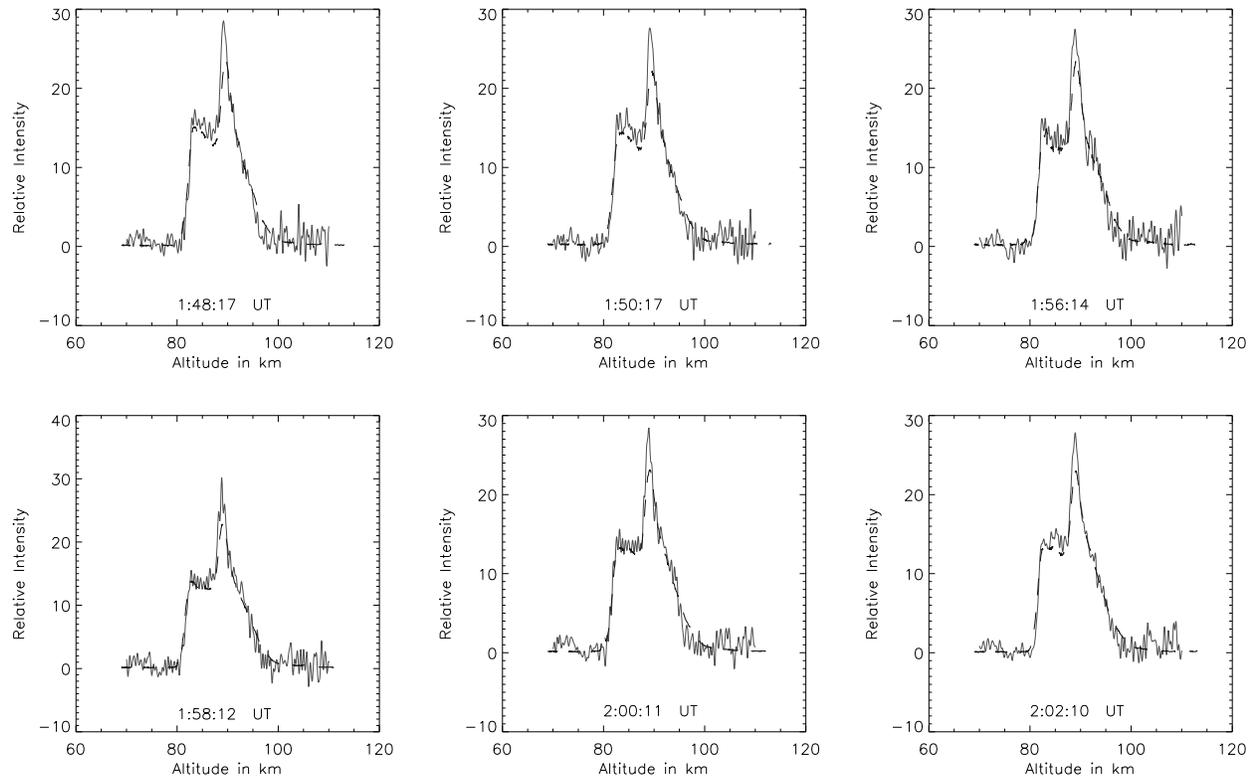}}
\caption{Results of the sodium profile on Oct 17 derived from the
LIDAR experiment (solid line), compared to the direct imaging results
(dashed line).}
\label{lid3}
\end{figure*}

Profiles were obtained by summing rows of 10 pixels perpendicular to the
altitude axis.
Fig~\ref{lid3} shows that a very good agreement can be found between
the data originating from the two different observing techniques.
In these data CCD imaging of the sodium layer appears to give  profiles with 
better signal-to-noise  than  the LIDAR technique,  but this is likely
to be due to a reduced count rate in the APD FoV either resulting from
the poor seeing or problems with alignment.   
The effect of the poor seeing is certainly apparent in the amplitude of
the sporadic layer:
it reduced the height resolution of the imaging data to approximately
1\,km, far inferior to the 150\,m attainable with LIDAR.
The discussion above also confirms that LIDAR is superior
in terms of absolute height calibration.

%---------------------------------------------------

\section{Expected performance of the VLT sodium profiler}
\label{expected_performanceVLT}

\subsection{Some considerations}
\label{some_consid}

The concept of the VLT sodium profiler (Butler et al. \cite{but02})
is the same as that built for ALFA. 
The differences are mostly in  hardware in the sense that 
the  profiler is more automated and will integrate
 well into a VLT instrument  control system. 
Additionally, the VLT profiler will have a longer 32k bit pulse sequence
mainly for lower cross-correlation noise.
Below, we summarize the salient points learned from the ALFA LIDAR 
experiment, and from signal-to-noise calculations for the VLT sodium
profiler.

Firstly, laser pulse length affects height resolution and signal-to-noise.
A shorter pulse length allows a higher resolution in the sodium layer
profile measurement: 1\,$\mu$s corresponds to 150\,m, 
2\,$\mu$s to 300\,m, etc. 
However, there are advantages to a longer pulse:
it gives better signal-to-noise in the profile and a more accurate mean
height measurement than can be achieved simply by binning the data
afterwards. It will be possible to adjust the pulse length in the VLT LGS
LIDAR system over the range 0.1-10\,$\mu$s. It is assumed that the
data is collected over time periods equal to half the pulse length.

Secondly, there is the issue of  noise. 
The basic profile recovered by LIDAR is the apparent distribution 
from which the photons originate, and in this
profile the noise is independent of height. 
To derive the sodium atom density
distribution, each point in this basic profile must be multiplied by the
square of its distance. 
As a result, the noise in the density profile
increases quadratically with distance.

Thirdly, we should ask what is the most sensible average height calculation. 
The most obvious method to find the average height of the sodium layer
is to calculate the centroid height of the atoms in the 
layer, as we did for the ALFA LIDAR experiment.
But this has an intrinsic bias and the value calculated actually depends
on the resolution used.
An alternative is to use the median height of the atoms,
which is independent of the resolution (at least with pulses of up to
4\,$\mu$s) as long as the profile changes only slowly around the
600\,m either side of the median height; 
linear interpolation can be used between these points. 
However, the WFS detects photons, and so instead the apparent
distribution from which the photons originate should be used.
This differs from the actual
density distribution: for a uniform sodium atom distribution, 25\%
fewer photons will be detected from a distance of       
95\,km than from 85\,km. 
Probably the best method is to find the height above which (and below
which) half the sodium-line photons originate. 
In this paper, it is called the `median flux height'.

Finally, there may be a systematic height offset.  
With the LIDAR system on ALFA, there was a large constant offset of
270\,m in the height measurement. 
This is likely also to occur with the VLT LGS
system, but can likewise easily be calibrated by performing a LIDAR
measurement with a  reflective layer (such as the observatory dome) at
almost zero height.

\subsection{Simulations}

\begin{table}
\begin{tabular}{ll}
sodium  column density          & $2 \times 10^{13}$\,m$^{-3}$ \\
atmospheric  transmission       & 75\%                         \\
laser power  (during  a  pulse) & 5\,W                         \\
pulse  length                   & 1\,$\mu$s                    \\
launch transmission             & 60\%                         \\
collector transmission          & 1\%                          \\
zenith distance                 & $0^\circ$ / $60^\circ$       \\
integration time                & 30\,s /  60\,s               \\
\end{tabular}
\caption{Parameters used in the simulations of the VLT profiler}
\label{tab:pars}
\end{table}

\begin{table}
\begin{tabular}{lll} 
\hline\\
Integration time & $\sigma_{\rm H}$ (z=0$^\circ$) & 
$\sigma_{\rm H}$ (z=60$^\circ$) \\
 (s) & (m) & (m) \\
\hline\\
    30 & 47 & 318 \\
    60 & 39 & 199 \\
\hline\\
\end{tabular}
\caption{Standard deviation, $\sigma_H$, for the median flux height for 30\,s
   and 60\,s integrations at zenith distances of 0$^\circ$ and
   60$^\circ$.  A pulse length of 1\,$\mu$s has been used.}
\label{sstddev_hgt}
\end{table}

\begin{figure}
\centerline{ \includegraphics[width=8.4cm]{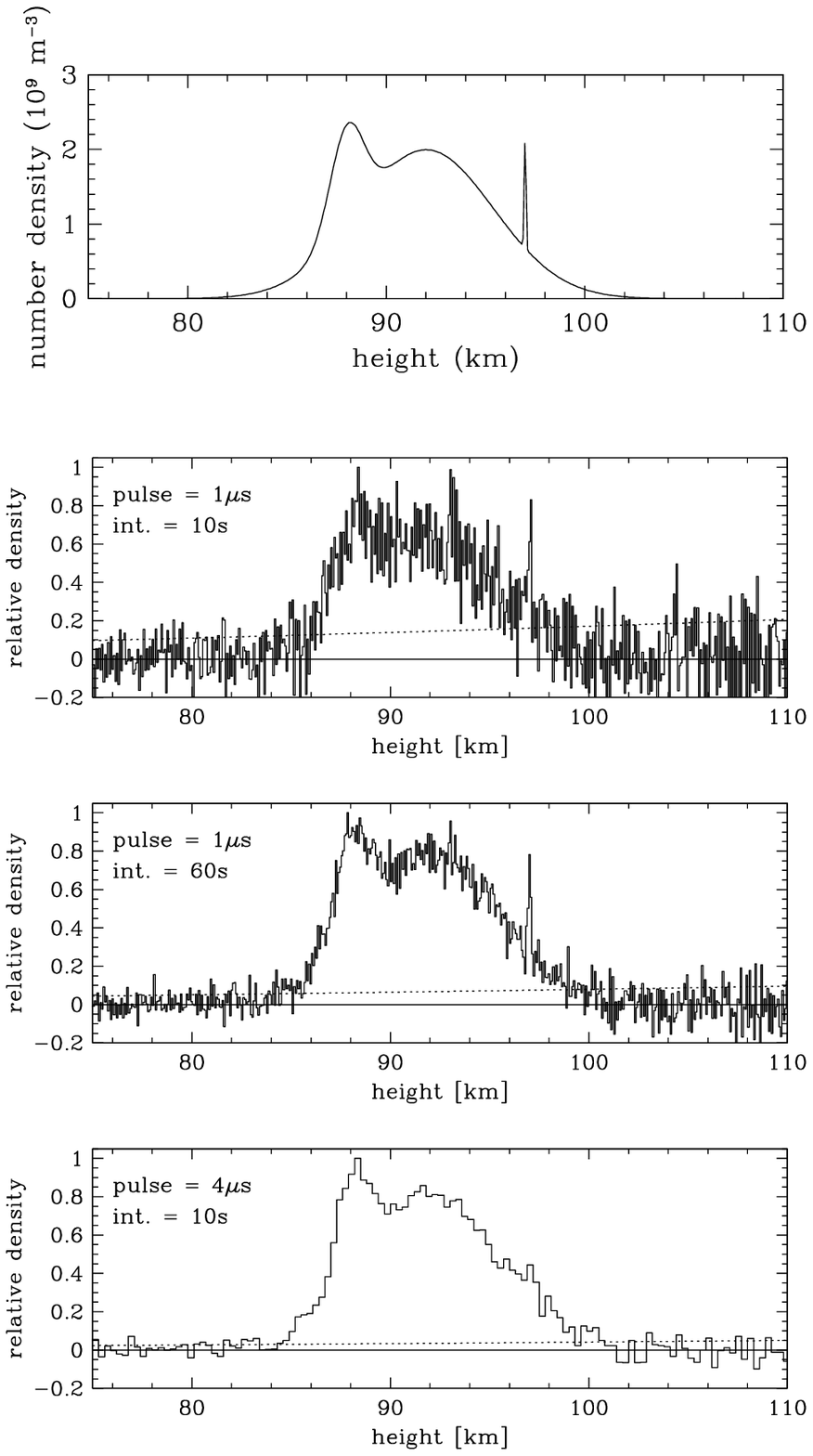}}
\caption{Top panel: The model sodium layer used in the LIDAR
simulations. 
Lower three panels:  Recovered profiles for different
   pulses widths and integration times.}  
\label{fig:model_Na_profile}
\end{figure} 

Simulations of LIDAR have been carried out to look at the
signal-to-noise issues and determine quantitatively what are
reasonable integration times, pulse lengths, etc. The uncertainties in
average height have been estimated from running 100 simulations for
each parameter set. 

The input to the model was the sodium density distribution shown in
the top panel of Fig.~\ref{fig:model_Na_profile}.
To indicate the difference in the `average' height obtained when it is
calculated in different ways, these are given for this profile as a
reference: 
91.52\,km (centroid height); 91.33\,km (median height); and 91.01\,km
(median flux height).
The difference between the centroid and median heights of the atoms is
200\,m; that between the median height of the atoms and the median height
from which the photons originate is an additional 300\,m. 
Since the acceptable error (Kissler-Patig, \cite{kis01}) is only
200\,m at zenith it is important to use the correct height estimator.

It has been assumed in the simulations that the APD will be mounted in one
of the AO systems at the VLT with an effective pupil diameter of
4.0\,m, which would provide a field of view of 3\arcsec. 
The collector transmission in the simulation is 1\% (which includes
the 2\% through a mirror in the AO system).
It has been assumed that transmission of the upward beam through the
LIDAR optics (in particular the modulator) is 50\%, so that
even during an `on' pulse the effective laser power is only 5\,W. This
is likely to be an underestimate.
The important parameters used in the simulation are given in
Table~\ref{tab:pars}.

The simulation is in two parts. 
One calculates the return flux
as a function of time measured by the MCS.
So that the data can be sampled twice for each pulse, each digit of the pulse
sequence is repeated (and the associated time-step halved).
The data for the sodium layer are then binned appropriately, and
convolved with the double-length pulse sequence.
The resulting vector is scaled so that it correctly reflects the
average flux over one sequence.
At this stage photon noise, background noise, and dark counts are
added.
This is repeated and the results co-added until the required
integration time is reached.
This represents the signal detected by the MCS.
The second part of the code takes this `observed' data and calculates
what the sodium profile was.
It is based on that used during the ALFA LIDAR experiment,
and is effectively a convolution of the pulse sequence (padded with an
extra zero after each digit) and the MCS data.
A background is then subtracted, the noise estimated, and the height
scaling imposed.

A set of three typical profiles are shown in the lower three panels of
Fig.~\ref{fig:model_Na_profile};
the central one of these corresponds to 
 the 60\,s integration time at zero zenith distance in
Table~\ref{sstddev_hgt}.
These demonstrate the effect on the profile of varying the
integration time and pulse length. 
In each of these, the dotted line marks the $3\sigma$ background noise
level.

Table~\ref{sstddev_hgt} shows the standard deviation (in metres) for the
median flux height, at zenith distances of 0$^\circ$ and 60$^\circ$,
and for 30\,s and 60\,s integrations. 
The 3-sigma uncertainty at zenith is
150\,m for a 30\,s integration, below the requirement of 200\,m. 
The simulations assume a relatively low sodium column density. 
Under conditions worse than this, it may not be possible to observe
using a LGS.
However, if a profile of the sodium layer were required, it would be
possible to reduce the error by as much as a factor two by
increasing the pulse length to 2--4\,$\mu$s.

\section{Conclusion}\label{conclusion_end}

\begin{enumerate}

\item 
We have developed and tested a method, based on LIDAR, of measuring
the height and profile of the mesospheric sodium layer using a cw
laser.
This avoids the need for a second telescope, has almost no dependence on the
atmospheric conditions, and has a very small error in the absolute
height.
The method is designed to be used at observatories with a LGS AO
system, and can be performed using the same laser during such
observations.

\item
We have tested our method with an experiment using the ALFA laser at
Calar Alto observatory in Spain.
The profiles derived from the LIDAR measurements compare well to those
obtained by simultaneous direct imaging and confirm that the method
can be used successfully.

\item  
The profiles we measured during this experiment demonstrate clearly that
there is a need to measure changes in sodium layer height in order to
set the initial wavefront sensor focus prior to, and maintain the
correct focus during, observations.
The difference in centroid height on consecutive nights was 1.5\,km,
and changes of several hundred metres were measured on timescales of a
few minutes.
The data also show large changes in the FWHM of the layer, which can have
consequences for the apparent elongation of the LGS spot as seen by a
wavefront sensor.

\item   
We have simulated the performance of a system to be integrated into
the VLT Laser Guide Star Facility.
This has highlighted the importance of choosing the correct method of
deriving the `average' distance to the sodium layer, and we propose
that the `median flux height' is the appropriate estimator.
A 3$\sigma$ error of less than 200\,m can be achieved with only a
30\,s integration under the required conditions.

\end{enumerate}

\begin{acknowledgements}
DB, RD, HF and NA acknowledge 
funding from the  TMR European Network for Laser Guide Stars at 8m Class Telescopes under contract ERBFMRXCT 960094.  DB acknowledges the support of the research and training network on
 `Adaptive Optics for Extremely Large Telescopes' under contract
 HPRN-CT-2000-00147. Jesus Aceituno,  Robert Weiss and the technical staff at MPIA 
 are thanked for valuable assistance during the experiment set-up.  S. Hippler
 is thanked for  help in preparing the colour pulse pattern figure.  
 It is a pleasure to thank the anonymous referee for a valuable comment. 
\end{acknowledgements}


\begin{thebibliography}{}
\bibitem[2000]{age00}
Ageorges N., Hubin N., 2000, \aaps, 144, 533

\bibitem[2002]{bon02}
 Bonaccini D., et al., 2002
 in {\em Adaptive Optical System Technologies II},
 eds. P. L. Wizinowich \& D. Bonaccini
 SPIE, 4839

\bibitem[2001]{bon01}
Bonnet H., 2001, 
in {\em SINFONI AO-Module: Control of the Trombone position in LGS
mode}, ESO document VLT-TRE-ESO-14710-2598, issue 1.0

\bibitem[2002]{but02} 
Butler D. J., Hippler S., Neumann U., et al.,  2002, 
 in {\em Adaptive Optical System Technologies II},
eds. P. L. Wizinowich \& D. Bonaccini
 SPIE, 4839

\bibitem[1982]{cle82}
Clemesha B.R., Simonich D.M., Batista P.P, Kirchhoff V.W.J.H., 1982, \jgr, 87, 181

\bibitem[1995]{cle95} Clemesha B.R., 1995,  J. Atmos. Phys., 57, 725

\bibitem[2002]{col02}
Collins S. C., Plane J. M. C., Kelley M. C., et al., 2002, 
J. Atmos. Sol. Terr. Phys., 64, 845

\bibitem[2000]{dav00}  
Davies R., Eckart A., Hackenberg W.,  et al., 2000, 
Expt. Ast., 10, 103

 \bibitem[2002]{dav02}
 Davies R.I., Ott T., Li J., Rabien S., Neumann U., Hippler S., 
 Bonaccini D., Hackenberg W., 2002,
 in {\em Adaptive Optical System Technologies II},
 eds. P. L. Wizinowich \& D. Bonaccini
 SPIE, 4839

\bibitem[2002]{org02}
D'Orgeville C., Rigaut F., Boccas M., et al., 2002, in {\em Adaptive Optical
  System Technologies II}, SPIE, 4839

\bibitem[1985]{foy85}
Foy R., Labeyrie, 1985,\aap, 152, L29

\bibitem[1991]{gardner91} Gardner C. S., Kane T. J., Hecht J. H,  et al.,
   1991, \grl,    18, 1369 

\bibitem[1998]{ge98}
Ge J., Jacobsen B. P., Angel J. R. P., McGuire P. C., Roberts T.,
McLeod B. A., Lloyd-Hart M., 1998,
in {\em Adaptive Optical System Technologies},
SPIE, 3353, eds. Bonaccini D., Tyson R. K., p.242

\bibitem[1993]{hec93}
Hecht J.H., Kane T.J., Walterscheid R.L., et al., 1993, J. Atmos. Terr. Phys., 55, 409

\bibitem[2000]{Kasper} Kasper M.,  Looze D., Hippler S., et al., 2000,  Exp. Ast.,  10, 49


\bibitem[2001]{kis01}
Kissler-Patig M., 2002
in {\em AO Top Level Requirements for the LGSF}, ESO document
VLT-SPE-ESO-11600-2369, issue 2.2

\bibitem[1988]{kwo88}
Kwon K.H., Senft D.C., Gardner C.S., 1988, \jgr, 93, 14199

\bibitem[2000]{lou00} 
Le Louarn M.,  2000, 
PhD thesis,  Univ. Lyon

\bibitem[1977]{meg77} 
Megie G., Bos F., Blamont. J. E., 1976, \planss, 225,  1093


\bibitem[2000]{mic00} 
Michaille L., Canas A. D., Dainty J. C., et al., 2000, \mnras, 139,  318
 

\bibitem[2001]{mic01}
Michaille L., Clifford, J. B., Dainty J. C., et al., 2001, 
 \mnras, 328,  993

\bibitem[2000]{sul00} 
O'Sullivan C. M. M., Redfern R. M., Ageorges N., et al.,  2000, 
Expt. Ast., 10, 147

\bibitem[2000]{ott00} 
Ott T., Hackenberg W., Rabien S., et al., 2000, 
Expt. Ast., 10, 89

\bibitem[1996]{pap96}
Papen G.C., Gardner C.S., Yu J., 1996, in {\em Adaptive Optics}, vol. 13, OSA
Technical Digest (OSA, Washington DC), 96.

\bibitem[2002]{pen02} Pennington D., et al., 2002,
 in {\em Adaptive Optical System Technologies II},
 eds. P. L. Wizinowich \& D. Bonaccini SPIE, 4839


\bibitem[2000]{rab00}  
Rabien S.,  Ott T.,  Hackenberg W.,  et al., 2000, Expt. Ast.,  10, 76


 \bibitem[2002]{rab02}
 Rabien S., Davies R.I., Ott T., Li J., Hippler S., Neumann U., 2002,
 in {\em Adaptive Optical System Technologies II},
 eds. P. L. Wizinowich \& D. Bonaccini
 SPIE, 4839

\bibitem[2000]{she00} 
She C. Y.,  Chen S.,  Hu Z.,  et al.,  2000, \grl, 27, 3289

\bibitem[2002]{sic02}
Sica R. J., Thayaparan T., Argall P. S.,  et al.,  2002, J. Atmos. Terr. Phys., 64, 915

\bibitem[1979]{sim79}
Simonich D.M., Clemesha B.R., Kirchhoff V.W.J.H., 1979, \jgr, 84, 1543
\end{thebibliography}
\end{document}